\documentclass[a4paper,11pt]{article}
\pdfoutput=1
\usepackage{jheppub}
\usepackage{bigints}
\usepackage{scalerel}
\usepackage[T1]{fontenc}

\newcommand\colequal{\mathrel{\overset{\makebox[0pt]{\mbox{\normalfont\tiny\sffamily col.}}}{=}}}

\title{Multipole Photon Radiation in the Vincia Parton Shower}

\author[a]{Peter Skands,}
\author[b,1]{Rob Verheyen\note{Corresponding author.}}

\affiliation[a]{School of Physics \& Astronomy, Monash University, Clayton VIC 3800, Australia}
\affiliation[b]{University College London, WC1E 6BT London, United Kingdom}

\emailAdd{r.verheyen@ucl.ac.uk}

\abstract{
We present algorithms that interleave photon radiation from the final state and the initial state with the QCD evolution in the antenna-based Vincia parton shower. 
One of the algorithms incorporates the complete soft and collinear structure associated with photon emission, but may be computationally expensive, while the other approximates the soft structure at a lower cost.
Radiation from fermions and $W$ bosons is included, and a strategy for photon radiation off leptons below the hadronization scale is set up.
We show results of the application of the shower algorithms to Drell-Yan and $W^+ W^-$ production at the LHC, showing the impact of the inclusion of the full soft structure and treatment of radiation off $W$ bosons.}

\begin{document}

\maketitle
\flushbottom

\section{Introduction} \label{introductionSection}
Parton-shower algorithms are an essential component of Monte Carlo event generators \cite{Review1}, providing a means of resumming radiative corrections in a fully exclusive and universal manner.
While the focus is typically on the simulation of QCD branching processes, the effects of the radiation of photons are in some cases substantial. 
For instance, photon radiation from the initial state and its interference with radiation from the final state has been shown to be significant for precision measurements at the LHC \cite{precision,precision2} and at future colliders \cite{future,future2, IFI1}.

Treatments of QED radiation based on collinear approximations are included in all of the standard general-purpose event-generator programs~\cite{Pythia8.2,Sherpa,Herwig++}. Modulo corrections from fixed-order process-specific QED matrix elements, 
these approaches neglect the eikonal interference structure.
On the other hand, YFS exponentiation \cite{yfs} is used in some cases \cite{decays1,decays2} as a means of including the soft interference structure in a universal process-independent way, but the QCD and QED showers are then not interleaved. In an interleaved evolution~\cite{Sjostrand:2004ef}, different branching types (here, QED and QCD ones)  are allowed to compete with each other for phase space during the shower evolution. This produces 
an arguably more physical relative ordering of evolution scales in the resulting joint resummation, compared to the non-interleaved case. 
In \cite{QEDantenna}, the first algorithm allowing QCD showers to be interleaved with a fully coherent (multipole) treatment of final-state QED 
radiation off fermions was set out, in the Vincia \cite{VinciaSimple,VinciaTimelike,VinciaHadron2} antenna-based parton-shower formalism. 
This letter describes the extension of that algorithm to initial-state radiation and photon radiation off $W$ bosons, 
and the implementation in the Vincia parton shower. 

The singularity structure associated with soft and collinear photon emission is recounted in Section \ref{singularitySection}. 
Next, two versions of our QED shower are presented in Section \ref{showerSection}: one that incorporates the full soft interference structure, and one that includes only dipole terms. The latter is computationally faster but does not account for multipole effects beyond the dipole level. 
A description of the treatment of photon radiation below the hadronization scale is also included, 
though without attempting to account for hadronic form factors.
Finally, section \ref{resultSection} shows the impact of the soft structure and treatment of radiation off $W$ bosons in the context of LHC processes. Some further details may be found in \cite{myThesis}.

\section{Photon Emission Singularity Structure} \label{singularitySection}
We first review the factorization properties of an $n+1$-particle matrix element that includes a photon.
In the soft limit \cite{photonFactorization}, the squared matrix element factorizes according to
\begin{equation} \label{softQED}
|M_{n+1}(\{p\},p_j)|^2 = -8 \pi \alpha \sum_{x,y}^n \sigma_x Q_x \sigma_y Q_y \frac{s_{xy}}{s_{xj}s_{yj}} |M_n(\{p\})|^2,
\end{equation}
where $\alpha$ is the fine-structure constant, $p_j$ is the momentum of the emitted photon and the sums run over all charged particles with momenta in the set $\{p\}$. 
The quantities $s_{xy} \equiv 2 p_x{\cdot}p_y$ indicate the Lorentz-invariance of the eikonal factor.
Note that when $x=y$, the inner product reduces to the invariant mass $2p_x{\cdot}p_x = 2 m_x^2$.
The factors $Q_x$ are the charges of particle $x$, while $\sigma_x = \pm 1$ is a sign factor that has $\sigma_x = 1$ for final-state particles and $\sigma_x = -1$ for initial-state particles. 
Charge conservation for the total event is then given by 
\begin{equation}
\sum_x Q_x \sigma_x = 0.
\end{equation}
The quasi-collinear limit \cite{QuasiCollinear1,QuasiCollinear2} with charged particle $i$ leads to
\begin{equation} \label{collinearQED}
|M_{n+1}(p_1,..,p_i,..,p_n,p_j)|^2 = 4\pi \alpha \, Q_i^2  \frac{2}{s_{ij}} P_{I \rightarrow ij}(z) |M_n(p_1,..,p_i+p_j,..,p_n)|^2,
\end{equation}
where the DGLAP splitting functions are given in terms of $z = E_i/E_{i+j}$ by 
\begin{align} \label{QEDsplittingFunctions}
P_{f\rightarrow f \gamma}(z) &= \frac{1 + z^2}{1-z} - 2 \frac{m_f^2}{s_{f \gamma}} \nonumber \\
P_{W^{\pm}\rightarrow W^{\pm} \gamma}(z) &= 2 \frac{z}{1-z} - 2 \frac{m_{\scaleto{W}{4pt}}^2}{s_{\scaleto{W}{4pt} \gamma}} + \frac{4}{3} \left( \frac{1-z}{z} + z(1-z) \right).
\end{align}
Note that the first two terms in the $W^{\pm}$ splitting function constitute the soft-collinear contribution that reduce to the eikonal factor in the soft limit.
The remaining pieces are purely collinear, and are weighted with a factor of $4/3$ because they are absent for the longitudinal $W$ polarization.
For initial-state radiation, the quasi-collinear limit is similar, with the exception of the inclusion of an additional factor $1/z$.

The branching kernels of the shower should capture the soft and quasi-collinear singularity structure of the matrix element factorization. 
In the QCD evolution of Vincia, they are incorporated in antenna functions spanned between pairs of partons.
In the photon emission case, the singularities may similarly be captured by the expression 
\begin{equation} \label{QEDemitAntennaSum}
a_{\mbox{\tiny{Emit}}}\left(\{p\}, p_j\right) = - \sum_{\{x,y\}} \sigma_x Q_x \sigma_y Q_y a_{\mbox{\tiny{Emit}}}(s_{xj}, s_{yj}, s_{xy}),
\end{equation}
where ${\{x,y\}}$ indicates that the sum runs over all \emph{pairs} of charged particles. 
The definition of the antenna-like functions $a_{\mbox{\tiny{Emit}}}(s_{xj}, s_{yj}, s_{xy})$ spanned between two charged particles depend on $x$ and $y$ being in the initial state or the final state.
They are given by 
\begin{align} \label{QEDemitAntennaeFF}
a_{\mbox{\tiny{Emit}}}^{\mbox{\tiny{FF}}}(s_{ij}, s_{jk}, s_{ik}) &= 4\frac{s_{ik}}{s_{ij} s_{jk}} - 4 \frac{m_i^2}{s_{ij}^2} 
    - 4 \frac{m_k^2}{s_{jk}^2} + \delta_{if} \frac{2}{s_{IK}} \frac{s_{jk}}{s_{ij}} + \delta_{kf} \frac{2}{s_{IK}} \frac{s_{ij}}{s_{jk}} \nonumber \\
&+ \delta_{iW} \frac{8}{3} \frac{1}{s_{ij}} \left(\frac{s_{jk}}{s_{IK} - s_{jk}} + \frac{s_{jk} (s_{IK} - s_{jk})}{s_{IK}^2} \right) \nonumber \\
&+ \delta_{kW} \frac{8}{3} \frac{1}{s_{jk}} \left(\frac{s_{ij}}{s_{IK} - s_{ij}} + \frac{s_{ij} (s_{IK} - s_{ij})}{s_{IK}^2} \right) \nonumber \\
a_{\mbox{\tiny{Emit}}}^{\mbox{\tiny{IF}}}(s_{aj}, s_{jk}, s_{ak}) &= 4 \frac{s_{ak}}{s_{aj} s_{jk}} - 4 \frac{m_a^2}{s_{aj}^2} - 4 \frac{m_k^2}{s_{jk}^2}
    + \delta_{af} \frac{2}{s_{AK}} \frac{s_{jk}}{s_{aj}} + \delta_{kf} \frac{2}{s_{AK}} \frac{s_{aj}}{s_{jk}} \nonumber \\
    &+ \delta_{aW} \frac{8}{3} \frac{1}{s_{aj}} \left( \frac{s_{jk}}{s_{AK} + s_{jk}} + \frac{s_{jk}}{s_{AK}} + \frac{s_{jk}^2}{s_{AK}^2} \right) \nonumber \\
    &+ \delta_{kW} \frac{8}{3} \frac{1}{s_{jk}} \left( \frac{s_{aj}}{s_{ak} + s_{jk}} +\frac{s_{aj}}{s_{AK} + s_{jk}} - \frac{s_{aj}^2}{(s_{AK} + s_{jk})^2} \right) \nonumber \\
a_{\mbox{\tiny{Emit}}}^{\mbox{\tiny{II}}}(s_{aj}, s_{bj}, s_{ab}) &= 4 \frac{s_{ab}}{s_{aj} s_{bj}} + \frac{2}{s_{AB}} \left(\frac{s_{aj}}{s_{bj}} + \frac{s_{bj}}{s_{aj}} \right),
\end{align}
where initial state particles are labelled by $a$ and $b$, final state particles are labelled by $i$, $j$, and $k$, and capital letters indicate the pre-branching momenta.
The Kronecker delta functions then check if the particle is a $W$ or a fermion. 
Note that they are absent from the initial-initial antenna, because $W$ bosons do not appear in hadronic or fermionic initial states.
They are however present in the initial-final antenna, because the $W$ may appear as the initial state of a resonance decay, which is showered as described in \cite{VinciaResonance}.

The parton shower approximation to the radiative matrix element is 
\begin{equation} \label{QEDemitPSApprox}
|M_{n+1}\left(\{p\}, p_j\right)|^2 \approx 4 \pi \alpha a_{\mbox{\tiny{Emit}}}\left(\{p\}, p_j\right) |M_n\left(\{\bar{p}\}\right)|^2,
\end{equation}
where $\{\bar{p}\}$ are the pre-branching momenta of the charged particles, related to the post-branching antennae by the kinematic maps \cite{VinciaMassive, VinciaHadron2}.
Considering the $s_{ij}$-collinear limit of eq.~\eqref{QEDemitAntennaSum} leads to 
\begin{align} \label{QEDemitAntennaSumCollinearLimit}
a_{\mbox{\tiny{Emit}}}\left(\{p\}, p_j\right) &= - \sigma_i Q_i \sum_{x \neq i} \sigma_x Q_x a_{\mbox{\tiny{Emit}}}(s_{ij}, s_{xj}, s_{ix}) + \mathcal{O}(1) \nonumber \\
&\colequal -  \sigma_i Q_i \frac{2}{s_{ij}} P_{I \rightarrow ij}(z) \sum_{x \neq i} \sigma_x Q_x \nonumber \\
&= Q_i^2 \frac{2}{s_{ij}} P_{I \rightarrow ij}(z).
\end{align}
As such, all collinear limits are automatically included in eq.~\eqref{QEDemitAntennaSum}. 
In a similar fashion, it may be shown that the mass terms in eq.~\eqref{softQED} are also properly incorporated. 
Finally, in the soft limit, each oppositely-charged particle pair (modulo the effects of crossing) contributes a positive term to the sum in eq.~\eqref{QEDemitAntennaSum} while each like-sign pair contributes a negative one; 
this produces the full pattern of constructive and destructive interference effects in the soft limit.
The negative contributions constitute a challenge from the Monte Carlo perspective, as they would preferably be incorporated without the introduction of negatively weighted events. 
The algorithms in the next section offer two different solutions. 

\section{Showering Algorithms} \label{showerSection}
In this section, we present two algorithms that are currently implemented in the Vincia parton shower. 
The first one captures the full soft structure indicated by eq.~\eqref{softQED}, but is at risk of becoming computationally expensive in certain situations. 
In the second algorithm, only the dominant dipole terms are kept in each phase-space point. 
This allows for a more efficient, and hence faster, algorithm, at the price of neglecting (subleading) corrections from quadrupole and higher multipole terms. 

\subsection{Coherent Algorithm} \label{coherentSubsection}
Our aim is to distribute emissions according to eq.~\eqref{QEDemitAntennaSum} while maintaining a structure that is as similar as possible to the QCD shower.
In particular, its ordering variable should regulate all singular limits simultaneously, but it should be directly comparable with the QCD ordering variable.
Furthermore, the kinematic mappings should be infrared safe, meaning that in all collinear limits all other charged particles should remain unaffected and in the soft limit no particle momenta should be modified. 

To meet the above requirements, we modify the parton shower approximation of eq.~\eqref{QEDemitPSApprox} to 
\begin{equation} \label{QEDemitPSApproxStep}
|M_{n+1}\left(\{p\}, p_j\right)|^2 \approx a_{\mbox{\tiny{Emit}}}\left(\{p\}, p_j\right) \sum_{\{x,y\}} \Theta\left(Q^2_{xy}\right) |M_n\left(\{\bar{p}\}_{xy}\right)|^2,
\end{equation}
where 
\begin{equation}
Q^2_{xy} = 
\begin{cases}
s_{xj} s_{yj}/s_{XY} &\mbox{ for final-final } \\
s_{xj} s_{yj}/(s_{XY} + s_{yj}) &\mbox{ for initial-final} \\
s_{xj} s_{yj}/s_{xy} &\mbox{ for initial-initial.}
\end{cases}
\end{equation}
is the transverse mometum in the antenna rest frame which is also used as the ordering scale in the QCD shower \cite{VinciaHadron2}.
In case of an initial-final antenna, $x$ here refers to the initial state while $y$ refers to the final state.
Eq.~\eqref{QEDemitPSApproxStep} then includes the step function
\begin{equation} \label{sectorStepFunction}
\Theta\left(Q^2_{xy}\right) =
\begin{cases}
1 & \text{if } \forall \text{ pairs } \{v,w\} \,\, Q^2_{xy} \leq Q^2_{vw} \\
0 & \text{otherwise.}
\end{cases}
\end{equation}
which divides the emissive phase space into sectors\footnote{See \cite{VinciaSector} for a sector-based approach to QCD antenna showers.}. 
Only a single term of the sum in eq.~\eqref{QEDemitPSApproxStep} thus contributes to each phase space point. 
For each such term, the argument of the non-radiative matrix element $\{\bar{p}\}_{xy}$ indicates that only the momenta $p_x$ and $p_y$ are modified; i.e., the recoil from the photon emission is shared by the pair of charged particles that has the lowest transverse momentum with the photon. 
Note that, while it contains negative contributions, the full kernel $a_{\mbox{\tiny{Emit}}}\left(\{p\}, p_j\right)$ is positive definite and may thus be generated without having to resort to negative weights.
Eq.~\eqref{QEDemitPSApproxStep} is thus relatively simple to implement in a shower using the usual Sudakov veto algorithm \cite{Fooling,Reloaded,Competing}. 
Competing trial emissions are generated in every sector using the appropriate local transverse momentum. 
An additional veto is included that checks the condition imposed by the step function in eq.~\eqref{sectorStepFunction}.

This procedure in fact orders emissions with ordering variable
\begin{equation} \label{orderingScaleMinimal}
Q^2 = \min\left(Q^2_{xy}\right),
\end{equation}
which has the required property of ensuring that all soft and collinear regions are contained in the limit $Q^2 \rightarrow 0$, while still allowing for the use of regular $2 \rightarrow 3$ shower kinematics.
However, this algorithm may become prohibitively expensive in situations where the number of charged particles in an event grows rapidly.

\subsection{Pairing Algorithm} \label{pairingSubsection}
To tackle the large computational cost of the above algorithm, the parton-shower approximation eq.~\eqref{QEDemitPSApprox} may instead be replaced by
\begin{equation} \label{QEDemitPSApproxPairing}
|M_{n+1}\left(\{p\}, p_j\right)|^2 \approx 4 \pi \alpha \sum_{[x,y]} Q^2_{[x,y]} a_{\mbox{\tiny{Emit}}}(s_{xj}, s_{yk}, s_{xy}) |M_n\left(\{\bar{p}\}_{xy}\right)|^2.
\end{equation}
The sum now runs over \emph{pairings} $[x,y]$ that have identical but opposite charge $Q_{[x,y]}$. 
Eq.~\eqref{QEDemitPSApproxPairing} trivially reduces to the correct collinear limits, but only contains a subset of eikonal factors.
By choosing a suitable method to pair up the charges, the missing interference structure may however be approximated.
\begin{figure}
\centering
\includegraphics[scale=0.7]{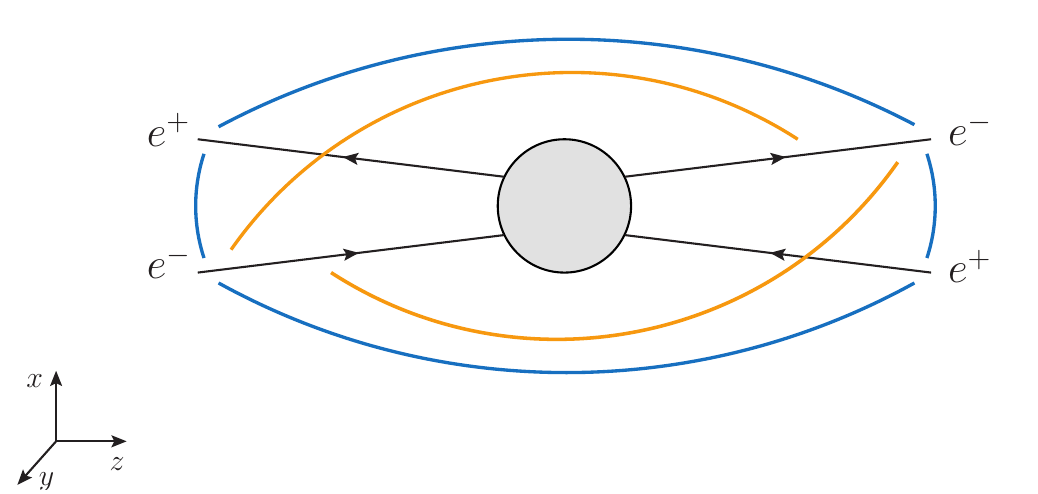}
\caption{An illustration of a $2e^+ 2e^-$ configuration where two pairs of nearby electron-positron are moving into roughly opposite directions. 
The blue lines indicate antennae with positive sign while the orange lines indicate antennae with negative signs.
In this scenario the contributions to the eikonal factor spanned between the pairs largely cancel, leaving only the positive contribution inside the pairs.}
\label{pairingFigure}
\end{figure}
To illustrate how this may be done, Figure \ref{pairingFigure} shows a configuration of charges consisting of two boosted $e^+ e^-$ pairs moving in opposite directions in space. 
In this situation, one pairing performs much better than the other.
Since the components of the pairs move in roughly the same direction, the charges of the electrons and positrons should be shielded and the radiation of photons should be suppressed.
Equivalently, contributions to eq.~\eqref{QEDemitPSApprox} between the pairs should largely cancel, and the remaining contributions are those inside the pair where the radiative phase space is restricted by the small invariant mass of the pair.
In this case, the soft structure would be mismodelled badly if the charges were paired up between pairs.
We therefore opt to pair up charges to minimize the sum of invariant masses of the pairs.
The combinatorial problem of finding the optimal pairing is known as the \emph{assignment problem}, which may be solved in $\mathcal{O}(n^3)$ time-complexity using the Hungarian algorithm \cite{hungarian1,hungarian2, hungarian3}.
Vincia makes used of an open-source implementation that may be found in \cite{hungarianImplementation}.
Note that it may not always be possible to pair up all charged particles with an opposite charged partner. 
For example, in a $W^+ \rightarrow u \bar{d}$ resonance decay, no pairing is possible at all. 
In these cases, the algorithm pairs up as many charges as possible, and employs the coherent algorithm on the remainder.

\subsection{Photon Emission Below the Hadronization Scale} \label{belowSubsection}
In the parton-shower formalism, a natural separation of scales occurs at $\Lambda_{\mbox{\tiny{QCD}}}$. 
Above that scale, coloured partons radiate gluons and multiple separate systems may radiate as a consequence of the simulation of resonance decays and multiple particle interactions in hadron collisions.
Below $\Lambda_{\mbox{\tiny{QCD}}}$, coloured partons hadronize and the evolution should continue as a single QED (multipole) system composed of leptons and charged hadrons, down to values beyond experimental precision. A detailed treatment of QED radiation off hadrons is beyond the scope of this work. However, even if we were to focus exclusively on radiation off leptons, the algorithm described in section \ref{coherentSubsection} is complicated by the fact that 
the system of leptons is not necessarily charge-conserving by itself.
We therefore continue the QED evolution below $\Lambda_{\mbox{\tiny{QCD}}}$ by using the algorithm of section \ref{pairingSubsection} and supplementing the pool of charges with the available colour-neutral strings that are entering the hadronization stage. Strings that have an overall electric charge can thereby act as recoilers for photon emission off leptons. Since we do not attempt to describe photon radiation off the strings themselves, we replace the antenna function in eq.~\eqref{QEDemitPSApproxPairing} by the final-state dipole function 
\begin{equation}
a^{\mbox{\tiny{FF}}}_{\mbox{\tiny{Dipole}}}(s_{ij}, s_{jk}, s_{ik}) = 4\frac{s_{ik}}{s_{ij}(s_{ij} + s_{jk})} 
- 4 \frac{m_i^2}{s_{ij}^2} + \frac{2}{s_{IK}}\frac{s_{jk}}{s_{ij}} 
\end{equation}
which only contains the soft and quasi-collinear singularity structure of the lepton $i$. 

\section{Results} \label{resultSection}
In this section, we apply the new QED shower algorithms implemented in Vincia to Drell-Yan and $W^+ W^-$ production at the LHC and investigate their differences. 
We also compare to the default (DGLAP-based) Pythia QED shower, whose dipole kinematics are based on a principle of ``maximum screening'' similar to that of our pairing algorithm.
The Pythia results are produced with Pythia 8.2 \cite{Pythia8.2} using the default tune and the NNPDF2.3 PDF sets \cite{NNPDF2.3}.
The Vincia results are produced using Vincia 2.3 \cite{VinciaHadron2} with Pythia 8.2, using the default tune and the same PDF set. 

\subsection{Drell-Yan}
It is not straightforward to disentangle the effects of soft photon coherence from other phenomena in LHC processes.
Here, we consider high invariant mass $pp \rightarrow e^+ e^-$ at centre-of-mass energy $\sqrt{s} = 14$ TeV with the cuts
\begin{equation}
m_{ee}^2 > 1 \text{ TeV, } p_{\perp,e} > 25 \text{ GeV and } |\eta_e| < 3.5
\end{equation}
on the leptons as well as the cuts
\begin{equation}
p_{\perp,\gamma} > 0.5 \text{ GeV and } |\eta_{\gamma}| < 3.5
\end{equation} 
on the photons.
As the hard scattering is always $q\bar{q} \rightarrow e^+ e^-$, the soft photon emission probability is affected by interference between the initial state and the final state. 
At invariant mass close to the $Z$ boson mass, this interference is suppressed by a factor of the order of the off-shellness of the $Z$ \cite{IFI1,IFI2}. 
This is a result of the relatively long-lived nature of the $Z$ boson close to its mass peak, causing the production and decay to remain separated.
However, at high invariant masses the $Z$ boson decays almost immediately and the interference spans the full emission spectrum.

\begin{figure}
\centering
\begin{minipage}{.5\textwidth}
    \centering
    \includegraphics[scale=0.62]{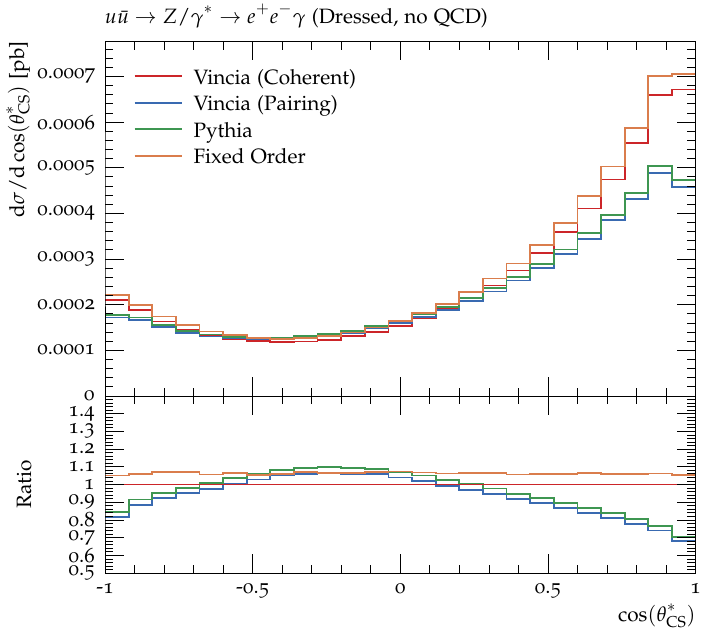}
\end{minipage}%
\begin{minipage}{.5\textwidth}
    \centering
    \includegraphics[scale=0.62]{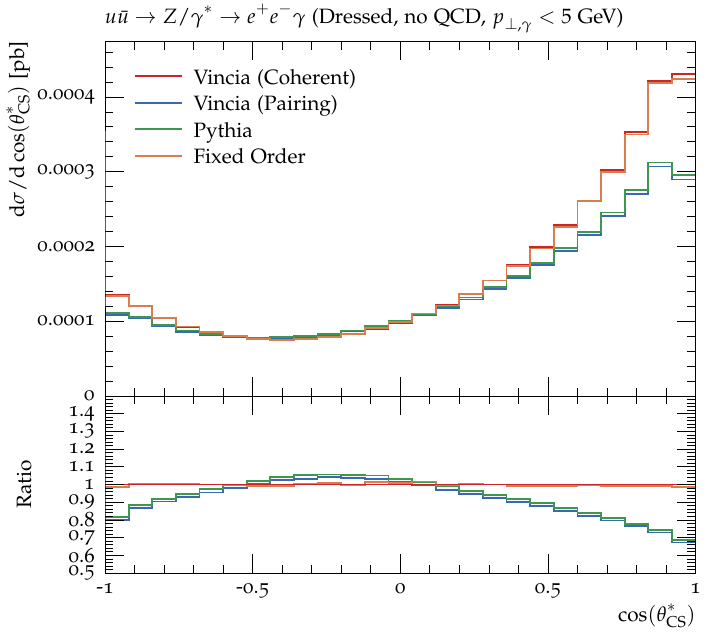}
\end{minipage}
\caption{Inclusive photon emission spectrum in $u \bar{u} \rightarrow e^+ e^-$ as a function of $\cos \theta_{\text{CS}}^*$ for the coherent (red) and pairing (blue) algorithms and the Pythia shower (green). 
Also included is the single-emission fixed order result of Madgraph \cite{Madgraph5} (yellow).
Shown are the fully inclusive spectrum (left) and the case where $p_{\perp, \gamma} < 5$ GeV is required (right).}
\label{DrellYanSterileFigure}
\end{figure}

\begin{figure}
\centering
\begin{minipage}{.5\textwidth}
    \centering
    \includegraphics[scale=0.62]{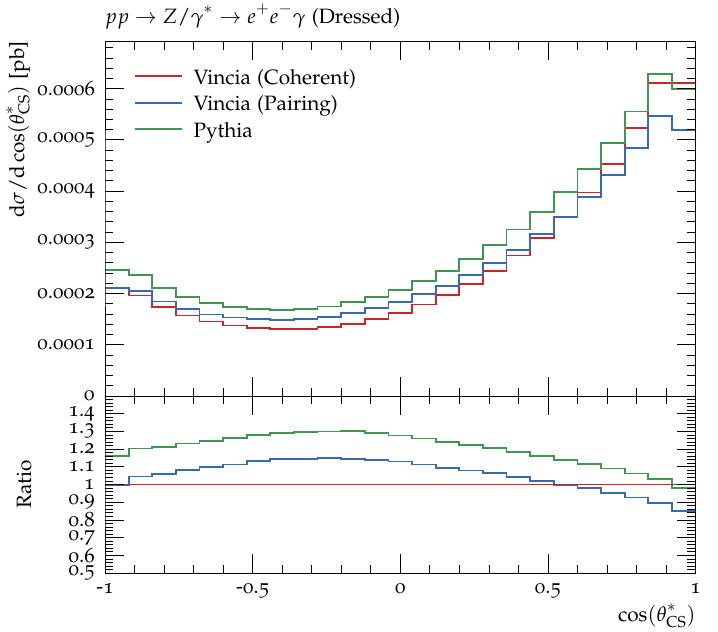}
\end{minipage}%
\begin{minipage}{.5\textwidth}
    \centering
    \includegraphics[scale=0.62]{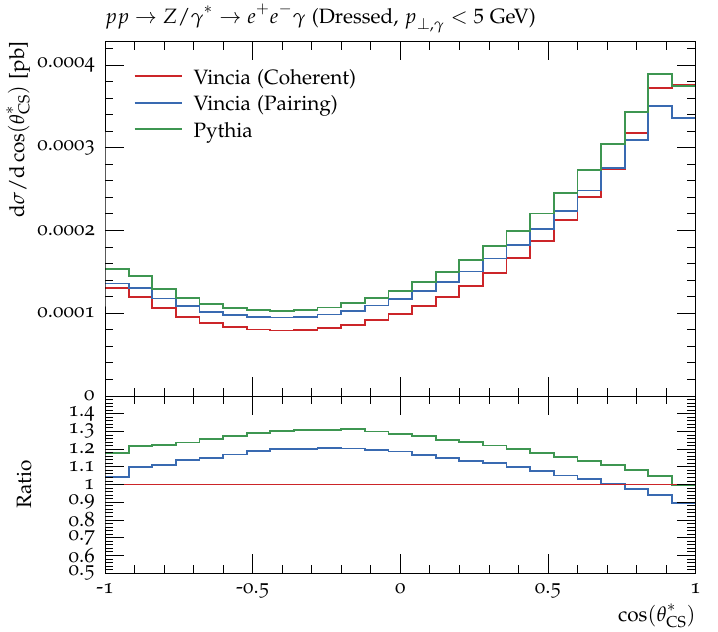}
\end{minipage}
\caption{Inclusive photon emission spectrum in $pp \rightarrow e^+ e^-$ as a function of $\cos \theta_{\text{CS}}^*$ for the coherent (red) and pairing (blue) algorithms and the Pythia shower (green). 
Shown are the fully inclusive spectrum (left) and the case where $p_{\perp, \gamma} < 5$ GeV is required (right).}
\label{DrellYanFigure}
\end{figure}

As a means of resolving the interference structure, we consider the photon emission probability as a function of the angle between the incoming quark and the outgoing electron in the Collins-Soper frame \cite{CSframe} where the impact of any unknown transverse momentum of the incoming (anti-)quark is minimized. 
However, the direction of the quark is ambiguous in $pp$ collisions. 
The angle is therefore defined with respect to the longitudinal boost of the $e^+ e^-$ pair, making the assumption that the initial state quark carries the largest momentum fraction. 
This assumption leads to a fraction of events where the quark direction is assigned incorrectly, but this fraction is relatively small as the anti-quark is always a sea quark and thus most often carries the least momentum.
The angle is then defined as
\begin{equation}
\cos \theta_{\text{CS}}^* = 2 \frac{p^z_{ee}}{|p^z_{ee}|} \frac{p_{e^+}^+ p_{e^-}^- - p_{e^+}^- p_{e^-}^+}{m_{ee} \sqrt{m_{ee}^2 + p_{\perp,ee}^2}},
\end{equation} 
where $p^{\pm} = (E \pm p^z)/\sqrt{2}$.
The influence of soft photon coherence is further muddled by QCD radiation and by the fact that the initial-state quarks may have positive or negative charges.
We first consider a more sterile environment where only the process $u \bar{u} \rightarrow e^+ e^-$ is included and QCD radiation is disabled. 
Figure \ref{DrellYanSterileFigure} shows the inclusive photon emission spectrum as a function of $\cos \theta_{\text{CS}}^*$ of the two Vincia algorithms and the Pythia shower. 
Also shown is the single-emission fixed order result as generated by Madgraph5 \cite{Madgraph5} with the same PDF set. 
The leptons are dressed by clustering them with photons within a cone distance $\Delta R = 0.2$.
On the left-hand side, all photons are included while on the right-hand side photons with $p_{\perp, \gamma} > 5$ GeV are excluded.
It is clear that only the coherent algorithm reproduces the fixed-order result, in particular in the soft limit here illustrated by the right-hand plot with $p_{\perp, \gamma}$.
Figure \ref{DrellYanFigure} shows the same distributions, but in the LHC process $pp \rightarrow e^+ e^-$ with QCD radiation enabled in the parton shower algorithms. 
As a consequence, differences between Pythia and Vincia appear but the shape difference of the coherent algorithms remains. 

\subsection{$W^+ W^-$ Production}
\begin{figure}
\centering
\begin{minipage}{.5\textwidth}
    \centering
    \includegraphics[scale=0.62]{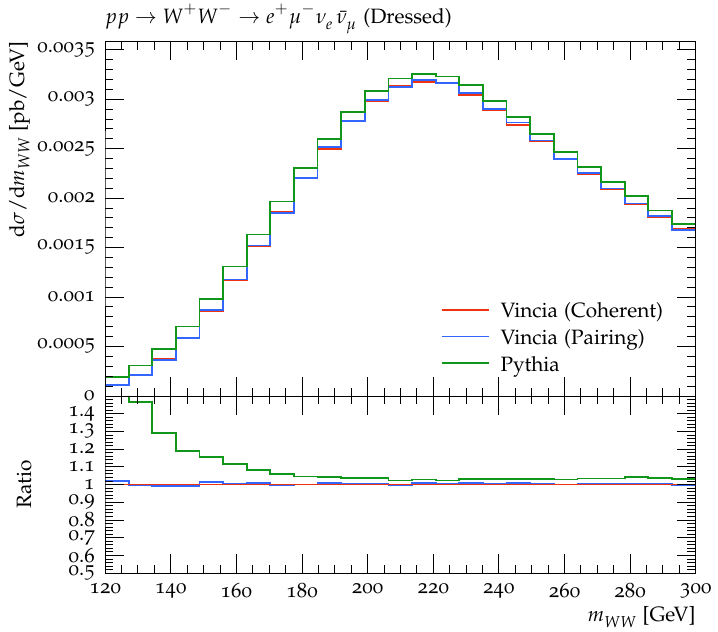}
\end{minipage}%
\begin{minipage}{.5\textwidth}
    \centering
    \includegraphics[scale=0.62]{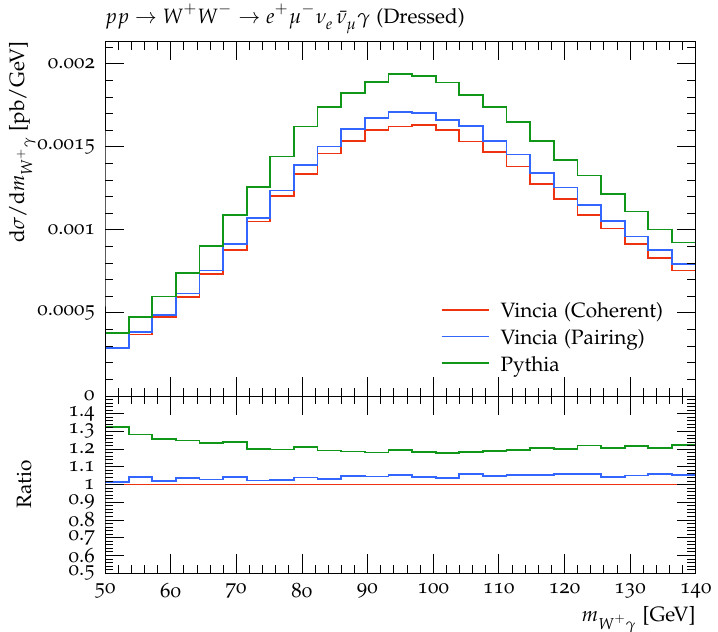}
\end{minipage}
\caption{Invariant mass spectra of the $W^+ W^-$ pair (left) and the $W^+$ with an isolated photon (right) for $pp\to W^+ W^-$ production at $\sqrt{s} = 14$ TeV using Vincia with the coherent algorithm (red), Vincia with the pairing algorithm (blue) and Pythia (green).
Made using RIVET~\cite{Rivet}.}
\label{WWFigure}
\end{figure}

We now consider invariant-mass observables for $pp\to W^+ W^-$ production with the leptonic decays $W^+ \rightarrow e^+ \nu_e$ and $W^- \rightarrow \mu^- \bar{\nu}_{\mu}$ at center-of-mass energy $\sqrt{s} = 14$ TeV.
The cuts 
\begin{equation}
p_{\perp} > 25 \text{ GeV and } |\eta| < 3.5
\end{equation}
are applied to the charged leptons, as well as the missing transverse energy cuts 
\begin{equation}
E_{\perp} > 20 \text{ GeV}
\end{equation}
applied to the neutrinos. 
Figure \ref{WWFigure} shows the spectra of the invariant mass of $W^+ W^-$ pair and the $W^+$ with an additional isolated photon. 
Invariant-mass distributions are good candidates to probe the effects of the QED shower because the QCD evolution of the initial state only affects the final state through recoil imparted by a Lorentz boost.
As such, invariant mass observables are unaffected by initial-state QCD radiation and isolate the QED corrections.
In this case, due to the more complex structure of the hard-scattering matrix element and the application of the phase space cuts, effects due to coherence are not visible.
However, differences between the Vincia showers and Pythia do appear, which are due to significant differences in the treatment of photon radiation off $W$ bosons and treatment of showers in resonance decays. 
The Pythia shower radiates photons from $W$ bosons using the fermionic DGLAP $f\to f\gamma$ splitting function, while the Vincia antenna functions include the full Yang-Mills coupling and the effects of the longitudinal $W$ boson polarization. 

\section{Conclusion}
In this letter, we have presented and implemented algorithms that interleave photon radiation in the Vincia parton shower, incorporating the full or approximate soft structure at varying computational cost.
The QED shower, which also includes photon splitting as described in \cite{QEDantenna} and radiation in resonance decays as described in \cite{VinciaResonance}, is available in Vincia 2.3 which has been incorporated into the Pythia event generator starting from Pythia version 8.3.
We have shown that the inclusion of interference effects between radiation from the initial state and final state affect the photon emission spectrum in Drell-Yan production at high invariant mass, 
and that the use of antenna functions for photon emission off $W$ bosons with the correct collinear limit leads to perceptible differences with the Pythia QED shower. 

\section{Acknowledgements}
We are grateful to Ronald Kleiss for many useful discussions. 
RV acknowledges support by the Foundation for Fundamental Research of Matter (FOM) via program 156 ”Higgs as Probe and Portal” and by the Science and Technology Facilities Council (STFC) via grant award ST/P000274/1. PS acknowledges support from the Australian Research Council via Discovery Project DP170100708. This work was also supported in part by the European Union's Horizon 2020 research and innovation programme under the Marie Sklodowska-Curie grant agreement No 722104 -- MCnetITN3.

\bibliographystyle{JHEP}
\bibliography{ref}

\end{document}